\documentclass
[prb,bibnotes,notitlepage,final,balancelastpage,superscriptaddress,twocolumn]{revtex4}%
\usepackage{amssymb}
\usepackage{amsmath}
\usepackage{amsfonts}
\usepackage{graphicx}
\usepackage{bm}
\usepackage{float}
\usepackage[bookmarks=false,pdfstartview=FitH,hyperindex=true, colorlinks,
linkcolor=blue, citecolor=blue]{hyperref}
\usepackage[all]{hypcap}%
\providecommand{\U}[1]{\protect\rule{.1in}{.1in}}

\begin{document}
\title{Quench dynamics of the topological quantum phase transition in the Wen-plaquette model}
\author{Long Zhang}
\affiliation{Hefei National Laboratory for Physical Sciences at
Microscale and Department of Modern Physics, University of Science
and Technology of China, Hefei, Anhui 230026, China}
\author{Su-Peng Kou}
\thanks{Corresponding author}
\email{spkou@bnu.edu.cn} \affiliation{Department of Physics, Beijing
Normal University, Beijing 100875, China}
\author{Youjin Deng}
\affiliation{Hefei National Laboratory for Physical Sciences at
Microscale and Department of Modern Physics, University of Science
and Technology of China, Hefei, Anhui 230026, China}

\begin{abstract}
We study the quench dynamics of the topological quantum phase
transition in the two-dimensional transverse Wen-plaquette model, which has a
phase transition from a $Z_{2}$ topologically ordered to a spin-polarized state.\ By
mapping the Wen-plaquette model onto a one-dimensional quantum Ising model, we
calculate the expectation value of the plaquette operator $F_{i}$ during a slowly
quenching process from a topologically ordered state.\ A logarithmic scaling law of quench dynamics
near the quantum phase transition is found, which is analogous to the well-known static
critical behavior of the specific heat in the one-dimensional quantum Ising model.

\end{abstract}
\maketitle

\section{Introduction}

Ultracold atoms provide an ideal platform for experimental studies of the time
evolution of quantum systems, and make it desirable for related theoretical
explorations on dynamics of quantum phase transitions in various
models.\ These explorations mainly focus on nonequilibrium dynamics in quantum
systems which undergo a quantum phase transition when a system parameter is varied
(\emph{quantum quench})\cite{Pol_colloquium}.\ These experimental and theoretical studies
can potentially help to pave the way for future technologies and provide a deeper understanding of
quantum many-body physics, particularly the universal scaling
behavior in the quench dynamics.

Recently, a new type of phase transition, the so-called topological quantum phase transition (TQPT) has
attracted considerable research attention\cite{Wenbook,wen3,wen4,wen1,Kitaev,nayak,lidar,chamon,xiang,yujing,vid1,vid2}%
.\ TQPT is a kind of phase transition between two quantum states with the
\emph{same} symmetry.\ It is fundamentally different from the usual
symmetry-breaking phase transition, and involves a new type
of order --- \emph{topological order}, as introduced by
Wen\cite{WenTo1}.\ In such an ordered state, there is no local order parameter,
and the state is robust against arbitrary local perturbations.\ On this basis,
quantum systems with topological order have been proposed to build robust
quantum memories\cite{Kitaev} and topological quantum computer
(TQC)\cite{kou1,kou2}. In Refs.\cite{kou1,kou2}, it was shown that the pure state of
topological order can be
obtained via an adiabatical and continuous evolution process from a
non-topologically ordered state and further be used as the initial state for TQC.\ Nevertheless,
the detailed dynamics of such a quench process, particularly the universal scaling behavior near TQPT,
has not been studied yet.

In the last decade, several exactly solvable spin models with topological order
were found, such as the toric-code model\cite{Kitaev}, the Wen-plaquette
model\cite{Wen} and the Kitaev model on a hexagonal lattice\cite{Kitaev2}.
These spin models provide a framework to study the TQPT and its
quench dynamics. Thus, in recent years, some research groups have studied the quench
dynamics of TQPT in the Kitaev model (e.g. Mondal~\emph{et al.}~\cite{quh-Kiv}) or the
toric-code model (e.g. Tsomokos~ \emph{et al.}~\cite{quh-tc}).\ Mondal
\emph{et al.} found a relationship between the quench rate and the defect
density in the one-dimensional (1D) and two-dimensional (2D) Kitaev model in the
limit of slow quench rate and generalized the result to the defect density of
a $d$ dimensional quantum model.\ Tsomokos \emph{et al.} investigated how a
topologically ordered ground state in the toric-code model reacts to rapid
quenches.\ They tested several cases and showed which kind of quench can
preserve or suppress the topological order.\

In this work, we study the quench
dynamics of TQPT from a topologically ordered state to a
non-topological order in the Wen-plaquette model.\ To characterize the phase transition, we
calculate the expectation value of plaquette operator $F_{i}$ during
the quenching process, which is related to the number of quasiparticles
in the topologically ordered state.\ Our results provide helpful information about
the whole quenching process.\

The remaining of this paper is organized as follows.\ Section II describes an exact mapping from the 2D transverse
Wen-plaquette model onto the 1D Ising chain.\ In Sec.\ III, we study the TQPT of
the 2D transverse Wen-plaquette model and give some results about its order
parameters. In Sec.\ IV, we give the solutions to the dynamics of TQPT in the
transverse Wen-plaquette model.\ A brief discussion is given in Sec.\ V.

\section{Mapping the transverse Wen-plaquette onto Ising model}

\begin{figure}[ptb]
\par
\begin{center}
\includegraphics[width=0.3\textwidth]{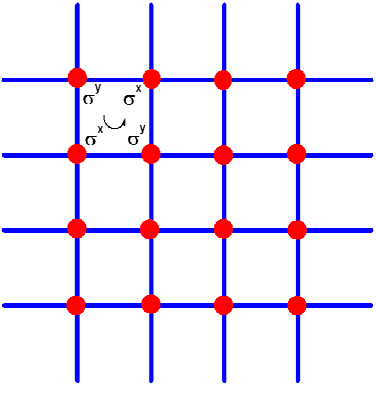}
\end{center}
\caption[9pt]{(Color online) The lattice where Wen-plaquette model is located.
A plaquette is defined by $F_{i}=\tau_{i}^{y}\tau_{i+\hat{x}}^{x}\tau
_{i+\hat{x}+\hat{y}}^{y}\tau_{i+\hat{y}}^{x}$.}%
\label{figure1}%
\end{figure}

We start with the Hamiltonian of the Wen-plaquette model
on a square lattice with periodic boundary conditions in both directions:
\begin{equation}
H_{W}=-g\sum_{i}F_{i},\quad F_{i}=\tau_{i}^{y}\tau_{i+\hat{x}}^{x}\tau
_{i+\hat{x}+\hat{y}}^{y}\tau_{i+\hat{y}}^{x}, \label{WenH0}%
\end{equation}
where $\tau_{i}^{x}$ and $\tau_{i}^{y}$ are Pauli operators on site $i$, $\hat{x}$ and
$\hat{y}$ are the unit vectors in x-axis and y-axis, respectively (see
Fig.~\ref{figure1}).\ Because of the commutativity of $H$ and $F_{i}$, the
energy eigenstates can be labeled by the eigenstates of $F_{i}$.\ We can easily
find $F_{i}^{2}=1$, so the eigenvalues of $F_{i}$ are $F_{i}=1$ and $F_{i}%
=-1$, which gives the exact ground state energy.\ In the case of $g>0$, the
ground state is $F_{i}=1$ for every plaquette, and the elementary excitation
is $F_{i}=-1$ on one plaquette (denoted by $i$) with an energy gap
$E_{g}-E_{0}=2g$.\ On an even-by-even lattice, there are two types of
plaquettes --- the even plaquettes and the odd plaquettes respectively. As a
result, one may define two kinds of bosonic quasiparticles: $Z_{2}$ charge and
$Z_{2}$ vortex (see detailed calculations in Ref. \cite{Wen} or Ref.
\cite{Wenbook}).\ A $Z_{2}$ charge is defined by $F_{i}=-1$ on an even
sub-plaquette while a $Z_{2}$ vortex defined by $F_{i}=-1$ on an odd. Thus a
fermion can be regarded as the bound state of a $Z_{2}$ charge and a $Z_{2}$ vortex.

Now we consider the Wen-plaquette model in a transverse field, which is
defined by:
\begin{equation}
H_{W}^{\prime}=-g\sum_{i}F_{i}-J\sum_{i}\tau_{i}^{x}.\label{WenH}%
\end{equation}
This model on a square lattice can be mapped onto the 1D quantum
Ising model with the Hamiltonian~\cite{yujing}
\begin{equation}
H_{I}=-\sum_{n=1}^{N}(g_I{\sigma}_{n}^{x}+{\sigma}_{n}^{z}{\sigma}_{n}%
^{z}),\label{eq:isingH}%
\end{equation}
where $\sigma_{n}^{x}$ and $\sigma_{n}^{z}$ are Pauli operators.

To derive the mapping, one can calculate the commutation relations (see detailed
calculations in Appendix): {\setlength\arraycolsep{1pt}
\begin{align}
&  [F_{i},\tau_{j}^{x}]=2F_{i}\tau_{j}^{x}(\delta_{i,j-\hat{x}}+\delta
_{i,j-\hat{y}}),\nonumber\label{commutation1}\\
&  [F_{i},F_{j}]=0,\nonumber\\
&  [\tau_{i}^{x},\tau_{j}^{x}]=0.
\end{align}
} These relations correspond to those in Ising model:
{\setlength\arraycolsep{1pt}
\begin{align}
&  [\sigma_{i}^{x},\sigma_{j}^{z}\sigma_{j+1}^{z}]=2\sigma_{i}^{x}\sigma
_{j}^{z}\sigma_{j+1}^{z}(\delta_{i,j}+\delta_{i,j+1}%
),\nonumber\label{commutation2}\\
&  [\sigma_{i}^{x},\sigma_{j}^{x}]=0,\nonumber\\
&  [\sigma_{i}^{z}\sigma_{i+1}^{z},\sigma_{j}^{z}\sigma_{j+1}^{z}]=0.
\end{align}
} Then we obtain the mapping
\begin{equation}
F_{i}\leftrightarrow\sigma_{i}^{x},\quad\tau_{i}^{x}\leftrightarrow\sigma
_{i}^{z}\sigma_{i+1}^{z}. \label{mapping}%
\end{equation}

Accordingly, the Hamiltonian (\ref{WenH}) can be mapped onto the 1D quantum Ising
model like following:
\begin{equation}
H_{W}^{\prime}\rightarrow-\sum_{\alpha}\sum_{i}(g\sigma_{\alpha i}^{x}%
+J\sigma_{\alpha i}^{z}\sigma_{\alpha i+1}^{z}),\label{WenH1}%
\end{equation}
where the subscript $\alpha$ implies there is one or more Ising chains.\ The
number of Ising chains is determined by the size of the square lattice for the
Wen-plaquette model (see Ref.\cite{yujing}).\ Since the Ising chains decouple from
each other, we can consider only one Ising
chain without loss of generality, and reduce the Hamiltonian (\ref{WenH1}) to
\begin{equation}
H_{W}^{\prime}=-J\sum_{i}(g_{W}\sigma_{i}^{x}+\sigma_{i}^{z}\sigma_{i+1}%
^{z}),\quad g_{W}=\frac{g}{J}.\label{WenH2}%
\end{equation}

Then we may explore the quantum properties of the original Wen-plaquette model
by studying the corresponding 1D Ising model.

\section{String order parameters in Wen-plaquette model}

For the 1D transverse Ising model (\ref{eq:isingH}), there are two phases: in the
limit of $g_{I}\gg1$, the ground state is a paramagnet with all spins
polarized along x-axis, $\left\langle \sigma_{i}^{x}\right\rangle
\rightarrow1$; in the limit of $g_{I}\ll1$, there are two degenerate
ferromagnetic ground states with all spins along positive or negative z-axis
and $\left\langle \sigma_{i}^{x}\right\rangle \rightarrow0$.\ Consequently, in
Hamiltonian (\ref{WenH2}), there is a quantum critical point at\cite{Baxter}
\begin{equation}
g_{W}=\frac{g}{J}=1
\end{equation}
that divides the two phases. Accordingly, the original transverse
Wen-plaquette model also has two phases separated by this quantum critical
point --- in the region of $g_{W}>1$, the system is a topologically ordered
state; in the region of $g_{W}<1,$ it's a spin-polarized state.

Noting that the local order parameters cannot be used to learn the nature of TQPT any
more, we introduce two non-local order parameters $\psi_{1}$ and $\psi_{2}$ as
string order parameters (SOP's) in the transverse Wen-plaquette model.\ They are
defined by the expectations of string operators $\prod_{i}F_{i}$ and
$\prod_{i}\tau_{i}^{x}$ with $i$ as the site index along a string in the diagonal direction\cite{yujing},
i.e. $\psi_{1}\equiv\left\langle \prod_{i}%
F_{i}\right\rangle $ and $\psi_{2}\equiv\left\langle \prod_{i}\tau_{i}%
^{x}\right\rangle $, respectively.

We then calculate these two SOP's by using the mapping in Eq.(\ref{mapping}).
For $\psi_{2}$, one has
\begin{equation}
\psi_{2}=\left\langle \prod_{i}\tau_{i}^{x}\right\rangle =\left\langle
\sigma_{1}^{z}\sigma_{2}^{z}\sigma_{2}^{z}\sigma_{3}^{z}\cdots\sigma_{n-1}%
^{z}\sigma_{n}^{z}\right\rangle =\left\langle \sigma_{1}^{z}\sigma_{n}
^{z}\right\rangle , \label{psi2_p}%
\end{equation}
which becomes the correlation of two spins in the Ising chain of length $n$.\ By employing
the Jordan-Wigner transformation
\begin{equation}
\sigma_{n}^{x}=1-2c_{n}^{\dagger}c_{n},\quad\sigma_{n}^{z}=-(c_{n}%
+c_{n}^{\dagger})\nu, \label{J_W_T}%
\end{equation}
where
\begin{equation}
\nu\equiv\prod_{m<n}(1-2c_{m}^{\dagger}c_{m})=\prod_{m<n}(c_{m}c_{m}^{\dagger
}-c_{m}^{\dagger}c_{m})=\prod_{m<n}A_{m}B_{m},
\end{equation}
where $A_{m}=c_{m}^{\dagger}+c_{m}$, $B_{m}=c_{m}^{\dagger}-c_{m}$, $c^\dagger_m$
and $c_m$ are the creation and annihilation operators for fermions,
{\setlength\arraycolsep{1pt} we obtain
\begin{align}
\psi_{2}  &  =\left\langle \sigma_{1}^{z}\sigma_{n}^{z}\right\rangle
=\left\langle (c_{1}+c_{1}^{\dagger})\nu(c_{n}+c_{n}^{\dagger})\right\rangle
\nonumber\label{psi2}\\
&  =\left\langle B_{1}A_{2}B_{2}\cdots B_{n-1}A_{n}\right\rangle .
\end{align}
Following the Wick's theorem, we can transform Eq.(\ref{psi2}) into a Toeplitz
determinant as\cite{book20}
\begin{equation}
\left\vert
\begin{array}
[c]{cccc}%
G_{12} & G_{13} & \cdots & G_{1n}\\
G_{22} & G_{23} & \cdots & G_{2n}\\
\vdots & \vdots & \ddots & \vdots\\
G_{n-1,2} & G_{n-1,3} & \cdots & G_{n-1,n}%
\end{array}
\right\vert , \label{to-det}%
\end{equation}
where
\begin{equation}
G_{ij}=-\frac{1}{2\pi}\int_{-\pi}^{\pi}\mathrm{d}k\frac{g_W-\cos k-i\sin
k}{\sqrt{(g_W-\cos k)^{2}+\sin^{2}k}}e^{ik(i-j)}. \label{G}%
\end{equation}
In the thermodynamic limit $N\rightarrow\infty$, we have (see Ref.
\cite{McCoy1} or Ref.\ \cite{McCoy2})
\begin{equation}
\psi_{2}\sim\left\{
\begin{array}
[c]{ll}%
(1-g_{W}^{2})^{1/4} & \text{when $g_{W}<1$}\\
0 & \text{when $g_{W}\geq1$}%
\end{array}
\right.  . \label{sol_psi2}%
\end{equation}
The exponent $1/4$ here agrees with the critical exponent $2\beta/\nu=1/4$ (see Ref.\cite{Baxter}).

By the same method we can obtain the result of $\psi_{1}$,%

\begin{equation}
\psi_{1}=\left\langle \prod_{i}F_{i}\right\rangle =\left\langle \prod
_{i}\sigma_{i}^{x}\right\rangle . \label{psi1}%
\end{equation}

In addition, we can take advantage of the duality of 1D Ising model (see
Ref.~\cite{book20}) :
\begin{equation}
s_{i}^{x}=\sigma_{i}^{z}\sigma_{i+1}^{z}\quad\text{and}\quad s_{i}^{z}%
=\prod_{k<i}\sigma_{k}^{x}. \label{dual}%
\end{equation}
Thus the result in Eq.(\ref{psi1}) is turned into $\psi_{1}=m$, where
$m\equiv\left\langle \sigma_{i}^{z}\right\rangle $ is the spontaneous
magnetization.\ We can find that when $N\rightarrow\infty$, the spin
correlation (\ref{psi2_p}) is the square of $m$. Consequently, from
Eq.(\ref{sol_psi2}), we have
\begin{equation}
\psi_{1}\sim\left\{
\begin{array}
[c]{ll}%
0 & \text{when $g_{W}\leq1$}\\
(1-g_{W}^{-2})^{1/8} & \text{when $g_{W}>1$}%
\end{array}
\right.  . \label{sol_psi1}%
\end{equation}

From the calculations above, one may notice that the non-local SOPs in a 2D
transverse Wen-plaquette model are transformed to the local order parameters in
the dual 1D Ising model.

\section{Quench dynamics of TQPT}

In this section we study the dynamics of TQPT in the transverse Wen-plaquette
model.\ The Kibble-Zurek mechanism (KZM)\cite{Kibble,Zurek} is a general
theory to explore the dynamics of second order phase transitions including the
quantum case.\ According to KZM, during a quench-induced phase transition, the
system undergoes three stages of evolution: adiabatic--pulse--adiabatic.\ It
predicts that the density of topological defects, which are generated by the
pulse evolution, is a function of quench time.

We can solve the dynamic problems in the Wen-plaquette model by taking a sequence
of transformations as in Ref.~\cite{quench}.\ First, through the
Jordan-Wigner transformation (\ref{J_W_T}), the spin operators $\sigma
_{i}^{x,z}$ are represented by fermionic operators $c_{n}$.\ Second, the operators $c_{n}$
are Fourier transformed into the momentum
space:
\begin{equation}
c_{n}=\frac{e^{-i\pi/4}}{\sqrt{N}}\sum_{k}c_{k}e^{ikn}. \label{Fourier}%
\end{equation}
Third, after Bogoliubov transformation, one have
\begin{equation}
c_{k}=u_{k}\eta_{k}+v_{-k}^{\ast}\eta_{-k}^{\dagger}, \label{Bogo}%
\end{equation}
where $\eta_{k}$ and $\eta_{k}^{\dagger}$ are fermionic operators.\ For
dynamic problems, the expression (\ref{Bogo}) becomes
\begin{equation}
c_{k}(t)=u_{k}(t)\widetilde{\eta}_{k}+v_{-k}^{\ast}(t)\widetilde{\eta}%
_{-k}^{\dagger}. \label{Bogo2}%
\end{equation}
Now, a dynamic solution can be written in the form of Bogoliubov mode
$(u_{k}(t),$ $v_{k}(t))$ after this whole transformation procedure.

By this method, the dynamics of the quantum Ising model can be expressed as the
time evolution of the Bogoliubov mode via so-called Bogoliubov-de Gennes
dynamic equations (see Ref.~\cite{quench}):
\begin{equation}
\left\{
\begin{array}
[c]{l}%
i\hbar\mathrm{d}u_{k}/\mathrm{d}t=+2(g_I(t)-\cos k)u_{k}+2\sin kv_{k}\\
i\hbar\mathrm{d}v_{k}/\mathrm{d}t=-2(g_I(t)-\cos k)v_{k}+2\sin ku_{k}%
\end{array}
\right.  . \label{eq:Bogo}%
\end{equation}

Consider a linear quench, i.e.
\begin{equation}
g_{I}(t<0)=-\frac{t}{{\tau}_{Q}}, \label{eq:g}%
\end{equation}
where $t$ varies from $-\infty$ to $0$ and the quench time $\tau_{Q}$
characterizes the quench rate (defined by $1/\tau_{Q}$).\ Equations
(\ref{eq:Bogo}) can be transformed into the form of Landau-Zener (LZ) model\cite{LZmodel}
(the connection between the KZM and the LZ model can be found in Ref.~\cite{Dam,Dam-Zuk}) :
\begin{equation}
\left\{
\begin{array}
[c]{l}%
i\hbar\mathrm{d}u_{k}/\mathrm{d}\tau=-\frac{1}{2}(\tau\Delta_{k}%
)u_{k}+\frac{1}{2}v_{k}\\
i\hbar\mathrm{d}v_{k}/\mathrm{d}\tau=+\frac{1}{2}(\tau\Delta_{k}%
)v_{k}+\frac{1}{2}u_{k}%
\end{array}
\right.  , \label{eq:LZM}%
\end{equation}
where
\[
\tau=4\tau_{Q}\sin k(\frac{t}{\tau_{Q}}+\cos k),
\]
and
\[
\Delta_{k}^{-1}=4\tau_{Q}\sin^{2}k.
\]

From equations (\ref{eq:LZM}), we derive a second order differential equation
for $v_{k}$:
\begin{equation}
\frac{\mathrm{d}^{2}v_{k}}{\mathrm{d}\tau^{2}}+(\frac{1}{4}\tau^{2}\Delta
^{2}_{k}+\frac{i\Delta_{k}}{2}+\frac{1}{4})v_{k}=0.
\end{equation}
After the substitutions
\begin{equation}
s=\frac{1}{4i\Delta_{k}}\quad\text{and}\quad z=\sqrt{\Delta_{k}}\tau
e^{i\pi/4},
\end{equation}
we have
\begin{equation}
\label{v_final}\frac{\mathrm{d}^{2}v_{k}(z)}{\mathrm{d}z^{2}}+(s+\frac{1}%
{2}-\frac{1}{4}z^{2})v_{k}(z)=0.
\end{equation}
This kind of differential equation has a general solution
\begin{equation}%
\begin{array}
[c]{l}%
v_{k}(\tau)=-[aD_{-s-1}(-iz)+bD_{-s-1}(iz)],\\
u_{k}(\tau)=(-\Delta_{k}\tau+2i\frac{\partial}{\partial\tau})v_{k}(\tau),
\end{array}
\label{general}%
\end{equation}
where $D_{m}(x)$ is the so-called parabolic cylinder function (PCF) or
Weber-Hermite function\cite{Whi-Wat}.

According to the boundary conditions and the characters of the PCF, one may
derive the approximative solutions to equations (\ref{v_final}) at the end of
linear quench for $t=0$ (See Ref.~\cite{quench}):
\begin{equation}%
\begin{array}
[c]{l}%
{|u_{k}|}^{2}=\frac{1-\cos k}{2}+e^{-2\pi\tau_{Q}\sin^{2}k},\\
{|v_{k}|}^{2}=1-{|u_{k}|}^{2},\\
u_{k}v_{k}^{\ast}=\frac{1}{2}\sin k+sgn(k)e^{-\pi\tau_{Q}\sin^{2}k}%
\sqrt{1-e^{-\pi\tau_{Q}\sin^{2}k}}e^{i\varphi_{k}},
\end{array}
\label{eq:IsingSolution}%
\end{equation}
with the condition $\tau_{Q}\gg1$.

\begin{figure}[ptb]
\par
\begin{center}
\includegraphics[width=0.45\textwidth]{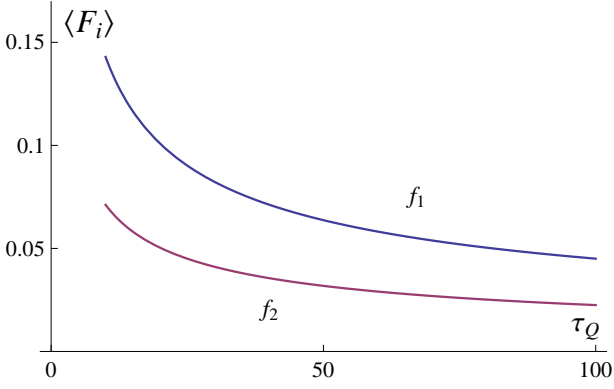}
\end{center}
\caption[9pt]{(Color online) The expectation value of $F_{i}$ at $t=0$ varying
with the quench time $\tau_{Q}$, where $f_{1}=1-\frac{1}{\pi}\int_{-\pi}^{\pi
}\mathrm{d}k(\frac{1+\cos k}{2}-e^{-2\pi\tau_{Q}\sin^{2}k})$ and the
approximative result $f_{2}=\frac{1}{\pi\sqrt{2\tau_{Q}}}$.}%
\label{figure2}%
\end{figure}

Now we calculate the dynamic solutions in the Wen-plaquette model via the same
transformation procedure plus the mapping (\ref{mapping}).\ We shall also consider
\begin{equation}
g_{W}(t<0)=-\frac{t}{{\tau}_{Q}}. \label{eq:g_I}%
\end{equation}
The quenching process can be set as tuning the strength of the transverse field from zero to very
large compared with the coupling constant $g$.\ The phase transition is thus
from a topologically ordered state to a spin-polarized state during the
time-evolution from $t\rightarrow-\infty$ to $t=0$.

First, we calculate the expectation value of $F_{i}$ after the quenching
process.\ From the Jordan-Wigner transformation, $\sigma_{n}^{x}%
=1-2c_{n}^{\dagger}c_{n}$, we have
\[
\left\langle F_{i}\right\rangle \rightarrow\left\langle \sigma_{n}%
^{x}\right\rangle =\left\langle \left(  1-2c_{n}^{\dagger}c_{n}\right)
\right\rangle .
\]
Through the Fourier transformation, we express $c_{m}^{\dagger}c_{n}$ in momentum
space
\begin{equation}
c_{m}^{\dagger}c_{n}=\frac{1}{2\pi}\int_{-\pi}^{\pi}\mathrm{d}k\int_{-\pi
}^{\pi}\mathrm{d}k^{\prime}c_{k}^{\dagger}c_{k^{\prime}}e^{i(k^{\prime}n-km)}.
\end{equation}
After the Bogoliubov transformation (\ref{Bogo}), we obtain\cite{intro1}
{\setlength\arraycolsep{1pt}
\begin{align}
\langle c_{m}^{\dagger}c_{n}\rangle &  =\frac{1}{2\pi}\int_{-\pi}^{\pi
}\mathrm{d}k\int_{-\pi}^{\pi}\mathrm{d}k^{\prime}e^{i(k^{\prime}n-km)}%
v_{-k}v_{-k}^{\ast}\delta_{k,k^{\prime}}\nonumber\\
&  =\frac{1}{2\pi}\int_{-\pi}^{\pi}\mathrm{d}k|v_{k}|^{2}e^{ik(n-m)}.
\end{align}
} From the solutions (\ref{eq:IsingSolution}), we derive
{\setlength\arraycolsep{1pt}
\begin{align}
\left\langle F_{i}\right\rangle  &  \rightarrow\left\langle \sigma_{n}%
^{x}\right\rangle =1-2\cdot\frac{1}{2\pi}\int_{-\pi}^{\pi}\mathrm{d}%
k(\frac{1+\cos k}{2}\nonumber\\
&  -e^{-2\pi\tau_{Q}\sin^{2}k})\overset{\tau_{Q}\gg1}{\simeq}\frac{1}{\pi
\sqrt{2\tau_{Q}}}.
\end{align}
}

We can see the original state in the limit of $g\gg1$, which is a topological
ground state with $\left\langle F_{i}\right\rangle \rightarrow\left\langle
\sigma_{n}^{x}\right\rangle \rightarrow1$, will finally evolve into the
trivial spin-polarized state with $\left\langle F_{i}\right\rangle
\rightarrow\left\langle \sigma_{n}^{x}\right\rangle \rightarrow0$%
.\ Figure~\ref{figure2} shows how the expectation value of $F_{i}$ at the end of
quenching process ($t=0$) varies with the quench rate.

We can obtain the the numbers of $Z_{2}$ charges and $Z_{2}$ vortices by
defining
\begin{equation}
\mathcal{N}_{c}\equiv\frac{1}{2}\sum_{i\in\mathrm{even}}(1-F_{i})=\sum
_{i\in\mathrm{even}}c_{i}^{\dagger}c_{i}%
\end{equation}
and%
\[
\quad\mathcal{N}_{v}\equiv\frac{1}{2}\sum_{i\in\mathrm{odd}}(1-F_{i}%
)=\sum_{i\in\mathrm{odd}}c_{i}^{\dagger}c_{i},
\]
respectively.\ Before the
quench, the system is in a topological ground state.\ There is no
quasiparticles ($N_{c}\equiv\langle\mathcal{N}_{c}\rangle=0$,
$N_{v}\equiv\langle\mathcal{N}_{v}\rangle=0$), so we have $F_{i}=1$
for all plaquettes.\ At the end of the quench, we can
also calculate the density of plaquettes $F_{i}=-1$\cite{intro2} by
\begin{align}
n=  &  \frac{1}{2N}\langle\sum_{i}(1-F_{i})\rangle=\frac{1}{2\pi}\int_{-\pi
}^{\pi}\mathrm{d}k(\frac{1+\cos k}{2}\nonumber\\
&  -e^{-2\pi\tau_{Q}\sin^{2}k})\overset{\tau_{Q}\gg1}{\simeq}\frac{1}{2}%
-\frac{1}{2\pi\sqrt{2\tau_{Q}}}.
\end{align}
It is obvious that when $\tau_{Q}\rightarrow\infty$, half of the plaquettes
will be turned into $F_{i}=-1$, which implies $\left\langle F_{i}\right\rangle
\rightarrow0.$

However, this result cannot give us any information about the quenching
process or the critical behaviors.\ In order to obtain such
information, we find that, for $\tau_{Q}\gg1$, the expression
\begin{equation}
|v_{k}(t)|^{2}=\frac{1}{2}(1+\frac{\cos k+t/\tau_{Q}}{\sqrt{1+2t/\tau\cos
k+(t/\tau_{Q})^{2}}})-e^{-2\pi\tau_{Q}\sin^{2}k}%
\end{equation}
is a time-dependent approximate function for the general solution
$|v_{k}(\tau)|^{2}$ in Eq.(\ref{general}), if we cut off the negative part of
the curve (see Fig.~\ref{figure3}).

\begin{figure}[ptb]
\par
\begin{center}
\includegraphics[width=0.45\textwidth]{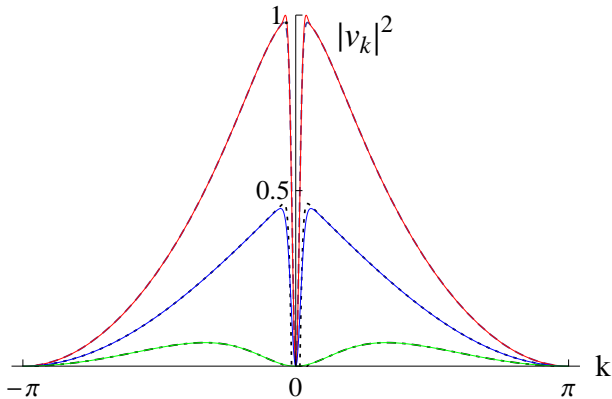}
\end{center}
\caption[9pt]{(Color online) The colored curves represent $\vert v_{k}%
(\tau)\vert^{2}$ in the original form of the PCF varying with momenta k from
$-\pi$ to $\pi$, where the red one is at $t=-0.5\tau_{Q}$, the blue
at $t=-\tau_{Q}$, the green at $t=-2\tau_{Q}$ with
$\tau_{Q}=50$ in all the three cases.\ The dashed, dotted and dot-dashed curves,
respectively, represent the approximate function for $\vert v_{k}(\tau
)\vert^{2}$ with corresponding parameters, with the negative parts under the
x-axis being cut off.\ They match with the colored curves very well}%
\label{figure3}%
\end{figure}

By using this approximate function, we calculate the value $\left\langle
F_{i}\right\rangle $ during the quenching process as a function of the time
$t$ near the quantum critical point.\ The result is shown in
Fig.~\ref{figure4}. \begin{figure}[ptb]
\par
\begin{center}
\hspace{60pt}
\includegraphics[width=0.42\textwidth]{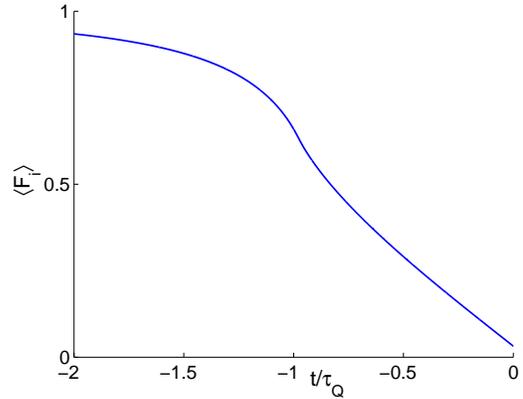}
\includegraphics[width=0.48\textwidth]{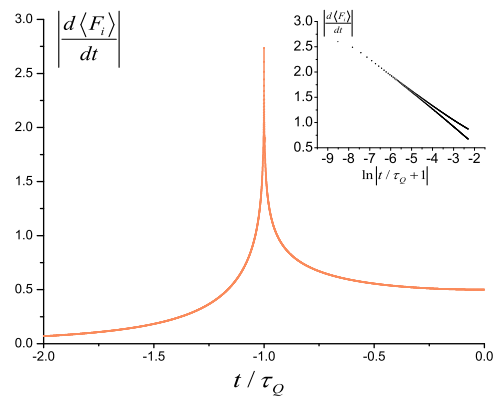}
\end{center}
\caption[9pt]{(Color online) The expectation value of $F_{i}$ during the
quenching process varying from $t=-2\tau_{Q}$ to $t=0$ with $\tau_{Q}=50$ and
the absolute value of the slope ($|\frac{\mathrm{d}\langle F_{i}\rangle}{\mathrm{d}t}|$)
corresponding to the curve above.\ The small graph shows
$|\frac{\mathrm{d}\langle F_{i}\rangle}{\mathrm{d}t}|\sim\ln|t/\tau_{Q}+1|$ from both
sides approaching the critical point $t=-\tau_{Q}$.}%
\label{figure4}%
\end{figure}Further calculations show that the derivative of the
expectation-value $\left\langle F_{i}\right\rangle $ diverges at point
$t=t_{c}=-\tau_{Q}$ as $\frac{\mathrm{d}\langle F_{i}\rangle}{\mathrm{d}%
t}\rightarrow-\infty$, which characterizes the critical point.\ In particular,
we obtain a logarithmic scaling law of quench dynamics near the quantum phase
transition as
\begin{equation}
\frac{\mathrm{d}\langle F_{i}\rangle}{\mathrm{d}t}\sim\ln|t-t_{c}|.
\end{equation}
Such a dynamics is analogous to the static scaling behavior of the specific heat near the
critical point for the 1D quantum Ising model.

The physical picture can be described as follows.\ At beginning, the external
field is weak and can be treated as a perturbation, which creates vortices or
charges and drives them to move around and annihilate each
other.\ As a result, the density of quasiparticles remains small.\ As the strength of
the field continuous to grow, the density of plaquettes $F_{i}=-1$ increases rapidly near the point
$g_{W}=1$ (that is $t=-\tau_{Q}$).\ The rate of density-changing diverges with
a logarithmic scaling law.\ Finally, at the end of the quench, the field
becomes so strong that the vortices and charges are all confined and cannot be
treated as quasiparticles.\ If the quenching process is infinitely slow, half of the plaquttes overturns.\

In addition we shall notice that only the linear quench is considered
here.\ However, for other cases, our method is not reliable.\ Inspired by
other papers on quantum quench in the toric-code model (e.g. Rahmani~\emph{et
al.}~\cite{Rah-Cha}, where a sudden quench is studied), we can study the
quench problems more generally through the time evolution of the entanglement
entropy in the Wen-plaquette model.\ Nevertheless, we will not consider this in this work.

\section{Conclusion}

In summary, we study the dynamics of TQPT caused by a linear quench in the transverse
Wen-plaquette model.\ We first show how to derive the mapping from the 2D
Wen-plaquette onto the 1D Ising model by comparing their commutation
relations.\ Based on this mapping, we point out the quantum critical point in
the transverse Wen-plaquette model and calculate its non-local order
parameters.\ We then calculate the expectation value of $F_{i}$ at the end of
the quench, and further show how this value varies during the whole
process.\ In particular, we find a logarithmic scaling law of quenching
process near the TQPT.

Finally we address the realization of the Wen-plaquette model in an optical
lattice of cold atoms.\ Because the Wen-plaquette model can be regarded as an
effective model of the Kitaev model on a two dimensional hexagonal lattice,
one may first realize the Kitaev model.\ The Hamiltonian of the Kitaev model
is\cite{Kitaev2}
\begin{equation}
\mathrm{H}=\sum_{j+l=\text{even}}(\mathrm{J}_{x}\sigma_{j,l}^{x}\sigma
_{j+1,l}^{x}+\mathrm{J}_{y}\sigma_{j-1,l}^{y}\sigma_{j,l}^{y}+\mathrm{J}%
_{z}\sigma_{j,l}^{z}\sigma_{j,l+1}^{z})
\end{equation}
where $j$ and $l$ denote the column and row indices of the lattice.\ In the
limit of $\mathrm{J}_{x}\mathrm{\gg J}_{z}\sim\mathrm{J}_{y}$ in this model,
the effective Hamiltonian of Kitaev model is simplified into that of the
Wen-plaquette model as
\begin{equation}
H_{0}=-\frac{\mathrm{J}_{z}^{2}\mathrm{J}_{y}^{2}}{16|\mathrm{J}_{x}|^{3}}%
\sum_{i}\sigma_{\text{left}(i)}^{x}\sigma_{\text{right}(i)}^{x}\sigma
_{\text{up}(i)}^{y}\sigma_{\text{down}(i)}^{y}. \label{fw}%
\end{equation}
Then one can use the Kitaev model in the limit $\mathrm{J}_{x}\mathrm{\gg
J}_{z}\sim\mathrm{J}_{y}$ on a torus to do the TQC.\ The realization of the
Kitaev model on the 2D hexagonal lattice has been proposed in
Ref.\cite{du,zo1}. The essential idea realizing the Kitaev model is to induce
and control virtual spin-dependent tunneling between neighboring atoms in the
lattice that results in a controllable Heisenberg exchange interaction.

\vskip0.5cm The authors acknowledge that this research is supported by NFSC
Grant No.\ 10874017,\ 10975127, National Basic Research Program of China (973
Program) under the grant No. 2011CB92180, the Anhui Provincial Natural Science
Foundation under Grant No.\ 090416224, and the Chinese Academy of Sciences.

\section*{Appendix}

In this part, we give detailed calculations about commutation relations
(\ref{commutation1}), which is related to the consistency between
Wen-plaquette model and Ising model.

The key point is to calculate $[F_{i},F_{j}]$, {\setlength\arraycolsep{1pt}
\begin{align}
&  [F_{i},F_{j}]\nonumber\label{apdx1}\\
&  =[\tau_{i}^{x}\tau_{i+\hat{x}}^{y}\tau_{i+\hat{y}+\hat{x}}^{x}\tau
_{i+\hat{y}}^{y},\tau_{j}^{x}\tau_{j+\hat{x}}^{y}\tau_{j+\hat{y}+\hat{x}}%
^{x}\tau_{j+\hat{y}}^{y}]\nonumber\\
&  =[\tau_{i}^{x},\tau_{j}^{x}\tau_{j+\hat{x}}^{y}\tau_{j+\hat{y}+\hat{x}}%
^{x}\tau_{j+\hat{y}}^{y}]\tau_{i+\hat{x}}^{y}\tau_{i+\hat{y}+\hat{x}}^{x}%
\tau_{i+\hat{y}}^{y}+\tau_{i}^{x}\nonumber\\
&  [\tau_{i+\hat{x}}^{y},\tau_{j}^{x}\tau_{j+\hat{x}}^{y}\tau_{j+\hat{y}%
+\hat{x}}^{x}\tau_{j+\hat{y}}^{y}]\tau_{i+\hat{y}+\hat{x}}^{x}\tau_{i+\hat{y}%
}^{y}+\tau_{i}^{x}\tau_{i+\hat{x}}^{y}\nonumber\\
&  [\tau_{i+\hat{y}+\hat{x}}^{x},\tau_{j}^{x}\tau_{j+\hat{x}}^{y}\tau
_{j+\hat{y}+\hat{x}}^{x}\tau_{j+\hat{y}}^{y}]\tau_{i+\hat{y}}^{y}+\tau_{i}%
^{x}\tau_{i+\hat{x}}^{y}\tau_{i+\hat{y}+\hat{x}}^{x}\nonumber\\
&  [\tau_{i+\hat{y}}^{y},\tau_{j}^{x}\tau_{j+\hat{x}}^{y}\tau_{j+\hat{y}%
+\hat{x}}^{x}\tau_{j+\hat{y}}^{y}].
\end{align}
} According to the commutation relations of Pauli operators,
{\setlength\arraycolsep{1pt}
\begin{align}
&  [\tau_{i}^{x},\tau_{j}^{x}\tau_{j+\hat{x}}^{y}\tau_{j+\hat{y}+\hat{x}}%
^{x}\tau_{j+\hat{y}}^{y}]\nonumber\\
&  =[\tau_{i}^{x},\tau_{j}^{x}]\tau_{j+\hat{x}}^{y}\tau_{j+\hat{y}+\hat{x}%
}^{x}\tau_{j+\hat{y}}^{y}+\tau_{j}^{x}[\tau_{i}^{x},\tau_{j+\hat{x}}^{y}%
]\tau_{j+\hat{y}+\hat{x}}^{x}\tau_{j+\hat{y}}^{y}\nonumber\\
&  +\tau_{j}^{x}\tau_{j+\hat{x}}^{y}[\tau_{i}^{x},\tau_{j+\hat{y}+\hat{x}}%
^{x}]\tau_{j+\hat{y}}^{y}+\tau_{j}^{x}\tau_{j+\hat{x}}^{y}\tau_{j+\hat{y}%
+\hat{x}}^{x}[\tau_{i}^{x},\tau_{j+\hat{y}}^{y}]\nonumber\\
&  =2i\delta_{i,j+\hat{x}}\tau_{j}^{x}\tau_{i}^{z}\tau_{j+\hat{y}+\hat{x}}%
^{x}\tau_{j+\hat{y}}^{y}+2i\delta_{i,j+\hat{y}}\tau_{j}^{x}\tau_{j+\hat{x}%
}^{y}\tau_{j+\hat{y}+\hat{x}}^{x}\tau_{i}^{z};\nonumber\\
&
\end{align}
} {\setlength\arraycolsep{1pt}
\begin{align}
&  [\tau_{i+\hat{x}}^{y},\tau_{j}^{x}\tau_{j+\hat{x}}^{y}\tau_{j+\hat{y}%
+\hat{x}}^{x}\tau_{j+\hat{y}}^{y}]\nonumber\\
&  =[\tau_{i+\hat{x}}^{y},\tau_{j}^{x}]\tau_{j+\hat{x}}^{y}\tau_{j+\hat
{y}+\hat{x}}^{x}\tau_{j+\hat{y}}^{y}+\tau_{j}^{x}[\tau_{i+\hat{x}}^{y}%
,\tau_{j+\hat{x}}^{y}]\tau_{j+\hat{y}+\hat{x}}^{x}\tau_{j+\hat{y}}%
^{y}\nonumber\\
&  +\tau_{j}^{x}\tau_{j+\hat{x}}^{y}[\tau_{i+\hat{x}}^{y},\tau_{j+\hat{y}%
+\hat{x}}^{x}]\tau_{j+\hat{y}}^{y}+\tau_{j}^{x}\tau_{j+\hat{x}}^{y}%
\tau_{j+\hat{y}+\hat{x}}^{x}[\tau_{i+\hat{x}}^{y},\tau_{j+\hat{y}}%
^{y}]\nonumber\\
&  =-2i\delta_{i+\hat{x},j}\tau_{j}^{z}\tau_{j+\hat{x}}^{y}\tau_{j+\hat
{y}+\hat{x}}^{x}\tau_{j+\hat{y}}^{y}-2i\delta_{i,j+\hat{y}}\tau_{j}^{x}%
\tau_{j+\hat{x}}^{y}\tau_{j+\hat{y}+\hat{x}}^{z}\nonumber\\
&  \tau_{j+\hat{y}}^{y};
\end{align}
} {\setlength\arraycolsep{1pt}
\begin{align}
&  [\tau_{i+\hat{x}+\hat{x}}^{x},\tau_{j}^{x}\tau_{j+\hat{x}}^{y}\tau
_{j+\hat{y}+\hat{x}}^{x}\tau_{j+\hat{y}}^{y}]\nonumber\\
&  =[\tau_{i+\hat{x}+\hat{x}}^{x},\tau_{j}^{x}]\tau_{j+\hat{x}}^{y}%
\tau_{j+\hat{y}+\hat{x}}^{x}\tau_{j+\hat{y}}^{y}+\tau_{j}^{x}[\tau_{i+\hat
{x}+\hat{x}}^{x},\tau_{j+\hat{x}}^{y}]\tau_{j+\hat{y}+\hat{x}}^{x}\nonumber\\
&  \tau_{j+\hat{y}}^{y}+\tau_{j}^{x}\tau_{j+\hat{x}}^{y}[\tau_{i+\hat{x}%
+\hat{x}}^{x},\tau_{j+\hat{y}+\hat{x}}^{x}]\tau_{j+\hat{y}}^{y}+\tau_{j}%
^{x}\tau_{j+\hat{x}}^{y}\tau_{j+\hat{y}+\hat{x}}^{x}\nonumber\\
&  [\tau_{i+\hat{x}+\hat{x}}^{x},\tau_{j+\hat{y}}^{y}]\nonumber\\
&  =2i\delta_{i+\hat{y},j}\tau_{j}^{x}\tau_{j+\hat{x}}^{z}\tau_{j+\hat{y}%
+\hat{x}}^{x}\tau_{j+\hat{y}}^{y}+2i\delta_{i+\hat{x},j}\tau_{j}^{x}%
\tau_{j+\hat{x}}^{y}\tau_{j+\hat{y}+\hat{x}}^{x}\nonumber\\
&  \tau_{j+\hat{y}}^{z};
\end{align}
} {\setlength\arraycolsep{1pt}
\begin{align}
&  [\tau_{i+\hat{y}}^{y},\tau_{j}^{x}\tau_{j+\hat{x}}^{y}\tau_{j+\hat{y}%
+\hat{x}}^{x}\tau_{j+\hat{y}}^{y}]\nonumber\\
&  =[\tau_{i+\hat{y}}^{y},\tau_{j}^{x}]\tau_{j+\hat{x}}^{y}\tau_{j+\hat
{y}+\hat{x}}^{x}\tau_{j+\hat{y}}^{y}+\tau_{j}^{x}[\tau_{i+\hat{y}}^{y}%
,\tau_{j+\hat{x}}^{y}]\tau_{j+\hat{y}+\hat{x}}^{x}\tau_{j+\hat{y}}%
^{y}\nonumber\\
&  +\tau_{j}^{x}\tau_{j+\hat{x}}^{y}[\tau_{i+\hat{y}}^{y},\tau_{j+\hat{y}%
+\hat{x}}^{x}]\tau_{j+\hat{y}}^{y}+\tau_{j}^{x}\tau_{j+\hat{x}}^{y}%
\tau_{j+\hat{y}+\hat{x}}^{x}[\tau_{i+\hat{y}}^{y},\tau_{j+\hat{y}}%
^{y}]\nonumber\\
&  =-2i\delta_{i+\hat{y},j}\tau_{j}^{z}\tau_{j+\hat{x}}^{y}\tau_{j+\hat
{y}+\hat{x}}^{x}\tau_{j+\hat{y}}^{y}-2i\delta_{i,j+\hat{x}}\tau_{j}^{x}%
\tau_{j+\hat{x}}^{y}\tau_{j+\hat{y}+\hat{x}}^{z}\nonumber\\
&  \tau_{j+\hat{y}}^{y},
\end{align}
} we can calculate (\ref{apdx1}) in following four cases:

\begin{enumerate}
\item when $i=j+\hat{x}$, {\setlength\arraycolsep{1pt}
\begin{align}
&  [F_{i},F_{j}]\nonumber\\
&  =2i\tau_{j}^{x}\tau_{j+\hat{x}}^{z}\tau_{j+\hat{y}+\hat{x}}^{x}\tau
_{j+\hat{y}}^{y}\tau_{i+\hat{x}}^{y}\tau_{i+\hat{y}+\hat{x}}^{x}\tau
_{i+\hat{y}}^{y}-2i\tau_{i}^{x}\tau_{i+\hat{x}}^{y}\tau_{i+\hat{y}+\hat{x}%
}^{x}\nonumber\\
&  \tau_{j}^{x}\tau_{j+\hat{x}}^{y}\tau_{j+\hat{y}+\hat{x}}^{z}\tau_{j+\hat
{y}}^{y}\nonumber\\
&  =-2\tau_{j}^{x}\tau_{j+\hat{x}}^{z}\tau_{j+\hat{y}+\hat{x}}^{z}\tau
_{j+\hat{y}}^{y}\tau_{i+\hat{x}}^{y}\tau_{i+\hat{y}+\hat{x}}^{x}+2\tau
_{i+\hat{x}}^{y}\tau_{i+\hat{y}+\hat{x}}^{x}\tau_{j}^{x}\tau_{j+\hat{x}}%
^{z}\nonumber\\
&  \tau_{j+\hat{y}+\hat{x}}^{z}\tau_{j+\hat{y}}^{y}\nonumber\\
&  =0;
\end{align}
}

\item when $i=j+\hat{y}$, {\setlength\arraycolsep{1pt}
\begin{align}
&  [F_{i},F_{j}]\nonumber\\
&  =2i\tau_{j}^{x}\tau_{j+\hat{x}}^{y}\tau_{j+\hat{y}+\hat{x}}^{x}\tau
_{j+\hat{y}}^{z}\tau_{i+\hat{x}}^{y}\tau_{i+\hat{y}+\hat{x}}^{x}\tau
_{i+\hat{y}}^{y}-2i\tau_{i}^{x}\tau_{j}^{x}\tau_{j+\hat{x}}^{y}\nonumber\\
&  \tau_{j+\hat{y}+\hat{x}}^{z}\tau_{j+\hat{y}}^{y}\tau_{i+\hat{y}+\hat{x}%
}^{x}\tau_{i+\hat{y}}^{y}\nonumber\\
&  =-2\tau_{j}^{x}\tau_{j+\hat{x}}^{y}\tau_{j+\hat{y}+\hat{x}}^{z}\tau
_{j+\hat{y}}^{z}\tau_{i+\hat{y}+\hat{x}}^{x}\tau_{i+\hat{y}}^{y}+2\tau
_{j+\hat{y}}^{z}\tau_{j}^{x}\tau_{j+\hat{x}}^{y}\tau_{j+\hat{y}+\hat{x}}%
^{z}\nonumber\\
&  \tau_{i+\hat{y}+\hat{x}}^{x}\tau_{i+\hat{y}}^{y}\nonumber\\
&  =0;
\end{align}
}

\item when $i+\hat{x}=j$ {\setlength\arraycolsep{1pt}
\begin{align}
&  [F_{i},F_{j}]\nonumber\\
&  =-2i\tau_{i}^{x}\tau_{j}^{z}\tau_{j+\hat{x}}^{y}\tau_{j+\hat{y}+\hat{x}%
}^{x}\tau_{j+\hat{y}}^{y}\tau_{i+\hat{y}+\hat{x}}^{x}\tau_{i+\hat{y}}%
^{y}+2i\tau_{i}^{x}\tau_{i+\hat{x}}^{y}\tau_{j}^{x}\tau_{j+\hat{x}}%
^{y}\nonumber\\
&  \tau_{j+\hat{y}+\hat{x}}^{x}\tau_{i+\hat{y}}^{z}\tau_{i+\hat{y}}%
^{y}\nonumber\\
&  =-2\tau_{i}^{x}\tau_{j}^{z}\tau_{j+\hat{x}}^{y}\tau_{j+\hat{y}+\hat{x}}%
^{x}\tau_{i+\hat{y}+\hat{x}}^{z}\tau_{i+\hat{y}}^{y}+2\tau_{i}^{x}\tau
_{i+\hat{x}}^{z}\tau_{j+\hat{x}}^{y}\tau_{j+\hat{y}+\hat{x}}^{x}\nonumber\\
&  \tau_{j+\hat{y}}^{z}\tau_{i+\hat{y}}^{y}\nonumber\\
&  =0;
\end{align}
}

\item when $i+\hat{y}=j$ {\setlength\arraycolsep{1pt}
\begin{align}
&  [F_{i},F_{j}]\nonumber\\
&  =2i\tau_{i}^{x}\tau_{i+\hat{x}}^{y}\tau_{j}^{x}\tau_{j+\hat{x}}^{z}%
\tau_{j+\hat{y}+\hat{x}}^{x}\tau_{j+\hat{y}}^{y}\tau_{i+\hat{y}}^{y}%
-2i\tau_{i}^{x}\tau_{i+\hat{x}}^{y}\tau_{i+\hat{y}+\hat{x}}^{x}\tau_{j}%
^{z}\nonumber\\
&  \tau_{j+\hat{x}}^{y}\tau_{i+\hat{y}+\hat{x}}^{x}\tau_{i+\hat{y}}%
^{y}\nonumber\\
&  =-2\tau_{i}^{x}\tau_{i+\hat{x}}^{y}\tau_{j}^{z}\tau_{j+\hat{x}}^{z}%
\tau_{j+\hat{y}+\hat{x}}^{x}\tau_{j+\hat{y}}^{y}+2\tau_{i}^{x}\tau_{i+\hat{x}%
}^{y}\tau_{i+\hat{y}+\hat{x}}^{z}\tau_{j}^{z}\tau_{j+\hat{y}+\hat{x}}%
^{x}\nonumber\\
&  \tau_{i+\hat{y}}^{y}\nonumber\\
&  =0.
\end{align}
}
\end{enumerate}

It is clear that we also have $[F_{i},F_{j}]=0$ in other cases.

Then, we only need to verify $[F_{i},\tau_{j}^{x}]=2F_{i}\tau_{j}^{x}%
(\delta_{i,j-\hat{x}}+\delta_{i,j-\hat{y}})$, while others can be easily
obtained from the commutation relations of Pauli operators. The verification
is shown in the following: {\setlength\arraycolsep{1pt}
\begin{align}
&  [F_{i},\tau_{j}^{x}]\nonumber\\
&  =-2i\delta_{j,i+\hat{x}}\tau_{i}^{x}\tau_{i+\hat{x}}^{z}\tau_{i+\hat
{y}+\hat{x}}^{x}\tau_{i+\hat{y}}^{y}-2i\delta_{j,i+\hat{y}}\tau_{i}^{x}%
\tau_{i+\hat{x}}^{y}\tau_{i+\hat{y}+\hat{x}}^{x}\tau_{i+\hat{y}}%
^{z}\nonumber\\
&  =-2\delta_{j,i+\hat{x}}\tau_{i}^{x}\tau_{i+\hat{x}}^{x}\tau_{i+\hat{x}}%
^{y}\tau_{i+\hat{y}+\hat{x}}^{x}\tau_{i+\hat{y}}^{y}-2\delta_{j,i+\hat{y}}%
\tau_{i}^{x}\tau_{i+\hat{x}}^{y}\tau_{i+\hat{y}+\hat{x}}^{x}\nonumber\\
&  \tau_{i+\hat{y}}^{x}\tau_{i+\hat{y}}^{y}\nonumber\\
&  =-2\delta_{j,i+\hat{x}}\tau_{j}^{x}F_{i}-2\delta_{j,i+\hat{y}}\tau_{j}%
^{x}F_{i}\nonumber\\
&  =2F_{i}\tau_{j}^{x}(\delta_{j,i+\hat{x}}+\delta_{j,i+\hat{y}}).
\end{align}
}


\begin{thebibliography}{99}
\bibitem {intro1}The expression (\ref{G}) is also calculated by this whole procedure.
\bibitem {intro2}It should be noticed that $Z_{2}$ charges and $Z_{2}$
vortices are on longer the elementary excitations or quasipartices of the
post-quench system.
\end{thebibliography}
\end{document}